\documentclass[prb,amsmath,amssymb,superscriptaddress, twocolumn ]{revtex4-1}
\usepackage{graphicx}
\usepackage{epsfig}
\usepackage{dcolumn}
\usepackage{bm}
\usepackage{rotating}
\usepackage{xspace}
\usepackage[table]{xcolor}
\usepackage[english]{babel}

\newcommand{\tio}{TiO$_2$\xspace}
\def\bk{{\bf k}}
\def\bq{{\bf q}}
\def\ve{\varepsilon}

\begin{document}
\widetext 

\title{Electron-plasmon and electron-phonon satellites \\[4pt] in the angle-resolved photoelectron spectra of $n$-doped anatase TiO$_2$}
\author{Fabio Caruso}
\affiliation{Institut f\"ur Physik and IRIS Adlershof, Humboldt-Universit\"at zu Berlin, Berlin, Germany}
\affiliation{Department of Materials, University of Oxford, Parks Road, Oxford, OX1 3PH}
\author{Carla Verdi}
\affiliation{Department of Materials, University of Oxford, Parks Road, Oxford, OX1 3PH}
\author{Samuel Ponc\'e}
\affiliation{Department of Materials, University of Oxford, Parks Road, Oxford, OX1 3PH}
\author{Feliciano Giustino}
\affiliation{Department of Materials, University of Oxford, Parks Road, Oxford, OX1 3PH}
\affiliation{Department of Materials Science and Engineering, Cornell University, Ithaca, New York 14853, USA}
\date{\today}

\begin{abstract}
We develop a first-principles approach based on many-body perturbation theory 
to investigate the effects of the interaction between electrons and carrier 
plasmons on the electronic properties of highly-doped semiconductors and oxides. 
Through the evaluation of the electron self-energy, we account
simultaneously for electron-plasmon and electron-phonon coupling 
in theoretical calculations of angle-resolved photoemission spectra, 
electron linewidths, and relaxation times. 
We apply this methodology to electron-doped anatase TiO$_2$ as an illustrative
example.
The simulated spectra indicate that electron-plasmon 
coupling in \tio underpins the formation of satellites at 
energies comparable to those of polaronic spectral features.
At variance with phonons, however, the energy of plasmons and their 
spectral fingerprints depends strongly on the carrier concentration, 
revealing a complex interplay between plasmon and phonon satellites. 
The electron-plasmon interaction accounts for approximately 40\% 
of the total electron-boson interaction strength and it is key 
to improve the agreement with measured quasiparticle spectra.
\end{abstract}
\maketitle

\section{Introduction}

The excitation of collective charge-density fluctuations\cite{pines1999elementary} (plasmons) 
is pervasive in solids and it manifests itself through distinctive 
signatures in spectroscopic experiments as, e.g., electron-energy 
loss spectroscopy (EELS) and integrated and angle-resolved photoemission 
spectroscopy (ARPES).
The emergence of plasmon satellites,\cite{Ferdi1996,guzzo/2011} for example, has been known 
since the early days of photoemission spectroscopy.\cite{Lundqvist1967b,Langreth1970,HEDIN19701,Baer1973}
More recently, the formation of plasmonic polaron bands in ARPES, 
dispersive spectral features that closely follow the
energy-momentum dispersion of the quasiparticle bands,
has been identified through first-principles calculations\cite{caruso/2015,caruso/2015/2,seidu2018} 
and confirmed experimentally in ARPES.\cite{Lischner/2015}

In $n$-type ($p$-type) doped semiconductors, the extrinsic 
carriers injected in the conduction (valence) band through 
the dopant atoms may host {\it carrier plasmons}, that is, low-energy 
plasmons with characteristic energies set by the carrier concentration $n$ 
via $\omega^{\rm pl} = (4\pi n /m_{\rm b}\epsilon_\infty)^{\frac{1}{2}}$, 
with $m_{\rm b}$ and $\epsilon_\infty$ being the band 
effective mass and the high-frequency dielectric constant, respectively.
Hartree atomic units are used throughout.
At variance with metals and undoped semiconductors, where the 
plasmon energy is of the order of 5-15~eV, for degenerate doping 
densities ($10^{17}$-$10^{19}$~cm$^{-3}$), the characteristic 
energy of carrier plasmons is typically smaller than 100~meV 
and it is thus comparable with the phonon energies of solids. 

At these energy scales, plasmons influence pervasively 
the electronic properties of doped semiconductors,  
in a similar fashion as the Fr\"ohlich interaction 
between electrons and longitudinal optical (LO) phonons in polar materials.\cite{Frohlich1954} 
First-principles calculations for highly-doped silicon, for example, illustrate that the 
scattering with plasmons induces the relaxation of excited quasi-particle 
states on sub-picosecond timescales, and a narrowing of the
band gap by 70~meV.\cite{PhysRevB.94.115208} These processes are forbidden
by energy conservation in the undoped parent compounds.
The emergence of low-energy photoemission satellites due to the 
coupling with carrier plasmons has further been observed in the 
photoemission spectrum of highly-doped oxides for both 
conduction bands\cite{EuO} and core levels\cite{SnO2}. 
These phenomena are reminiscent of the Fr\"ohlich interaction in polar
materials,\cite{Verdi2015,Verdi2017} revealing a striking similarity that calls for an 
in-depth comparison of plasmonic and polaronic coupling.

Highly-doped oxides constitute an optimal playground 
to explore the interplay between electrons, plasmons, and phonons as they 
simultaneously host polar LO phonon modes and tunable 
carrier plasmons, which lead to a rich scenario of electron-phonon 
and electron-plasmon coupling phenomena. 
In particular, the anatase phase of $n$-doped 
titanium dioxide (TiO$_2$) -- one of the most used 
materials in photocatalysis\cite{Kavan1996} and photovoltaic due to its stability, 
non-toxicity and natural abundance~\cite{DeAngelis2014} -- exhibits distinctive 
signatures of polaronic coupling, which manifest themselves through the 
emergence of photoemission satellites at binding energies around
100~meV below the conduction-band bottom.\cite{Moser2013prl,Moser2015}
The electronic properties of this polymorph have been investigated
in great detail from
first-principles calculations,\cite{Zhang2005,Deak2011,Mattioli2010,Kang2010,Chiodo2010}
making \tio an ideal candidate to explore the influence of the 
electron-plasmon interaction. 

In this work, we study the combined effect of plasmons and 
phonons on the electronic properties of highly-doped \tio. 
We develop a many-body approach to compute 
the electron self-energy due to the electron-plasmon 
interaction from first principles. 
At variance with earlier developments,\cite{PhysRevB.94.115208} this approach 
circumvents the high computational cost entailed by the explicit evaluation 
of the dielectric function in the presence of doping.  
By combining state-of-the-art calculations of the electron-phonon 
and electron-plasmon self-energy with the cumulant expansion 
approach,\cite{Gumhalter/2016/PRB} we investigate the influence of plasmons and phonons 
on the emergence of spectral signatures of bosonic coupling in ARPES.

Our first-principles calculations of ARPES spectra reveal that 
the coupling to plasmons may underpin the formation of satellites 
with energy and intensity comparable to those of polaronic satellites. 
Owing to the pronounced dependence of the plasmon energy on 
carrier concentration, the binding energy of plasmon satellites is 
modulated by doping in a range between 30 and 80~meV. 
While at doping concentrations lower than 10$^{19}$~cm$^{-3}$ 
the signatures of plasmon satellites may be hardly distinguishable from 
the quasiparticle bands owing to finite resolution effects, 
at higher carrier densities plasmons and phonons simultaneously contribute 
to the spectral intensity of photoemission satellites. 
Overall, our results indicate that the electron-plasmon 
interaction accounts for about 40\% of the total electron-boson 
interaction strength, which is corroborated by the improved 
agreement with the experimental quasiparticle weight. 
Calculations of the electron linewidths further indicate that the 
scattering rate of excited carriers is dominated by phonons, 
while the contribution of electron-plasmon scattering 
is of the order of 10-15\%.

The manuscript is organized as follows. In Sec.~\ref{sec:th} 
we review the many-body theory of electron-plasmon and electron-phonon 
coupling and the cumulant expansion approach. Details on the numerical 
evaluation of the electron-plasmon self-energy are given in Sec.~\ref{sec:imp}. 
The influence of plasmons and phonons on the photoemission spectrum of \tio is discussed in 
Sec.~\ref{sec:spec}, whereas linewidths and relaxation times are  
addressed in Sec.~\ref{sec:LW}. Finally, summary and concluding 
remarks are reported in Sec.~\ref{sec:conc}. 

\section{Theoretical background } \label{sec:th}
The interaction between electrons and bosonic excitations is commonly described within the formalism of many-body
perturbation theory through the electron-boson coupling self-energy in the Fan-Migdal approximation:\cite{Giustino2017}
  \begin{align}\label{eq:sigma}
  &\Sigma^{\rm e-b}_{n{\bf k}} (\omega) = \int\!\frac{d{\bf q}}{\Omega_{\rm BZ}}\sum_{m\nu}
     |g^{\rm e-b}_{mn\nu}({\bf k},{\bf q})|^2 \\
& \times  \left[ \frac { n_{{\bf q}\nu} + f_{m{\bf k+q}} } 
  {\omega - \varepsilon_{m{\bf k+q}} + \omega^{\rm b}_{{\bf q}\nu} - i\eta } 
  + \frac { n_{{\bf q}\nu} + 1 - f_{m{\bf k+q}} }
  {\omega - \varepsilon_{m{\bf k+q}} - \omega^{\rm b}_{{\bf q}\nu} - i\eta  } \right].\nonumber
  \end{align}
Here, { $g^{\rm e-b}$ denotes the electron-boson coupling matrix elements,}
$\omega^{\rm b}_{{\bf q}\nu}$ is the boson energy, ${\bf k}$ and ${\bf q}$ 
are Bloch wavevectors, $m$ and $n$ band indices, $\nu$ the branch index, $\varepsilon_{m{\bf k+q}}$ a 
set of single-particle eigenvalues, $n_{\bf q}$ and $f_{m{\bf k+q}}$ Bose-Einstein
and Fermi-Dirac occupations, respectively, and $\eta$  a positive infinitesimal. 
The summation runs over all Bloch states and the integral is over 
the Brillouin zone of volume $\Omega_{\rm BZ}$.
Equation~\eqref{eq:sigma} stems from the first-order expansion of the 
self-energy for a coupled electron-boson system interacting through the
Hamiltonian\cite{Mahan2000} {$\hat {H}^{\rm e-b} = \sum_{n m \nu} \sum_{\bk,\bq} \, g^{\rm e-b}_{nm\nu}(\bk,\bq)
\hat c_{m\bk+\bq}^\dagger \hat c_{n\bk} (\hat b_{\bq\nu} + \hat b^{\dagger}_{-\bq\nu})$}, with 
$\hat c$,$\hat c ^{\dagger}$ ($\hat b$,$\hat b ^{\dagger}$) 
fermionic (bosonic) annihilation and creation operators, respectively.
This Hamiltonian is commonly employed to describe electronic coupling to  
bosonic modes that can approximately be represented 
as a set of uncoupled harmonic oscillators as, for instance, 
phonons and plasmons.

The electron self-energy due to electron-phonon coupling\cite{grimvall1981electron}
is obtained from Eq.~\eqref{eq:sigma} by identifying $\omega^{\rm b}_{{\bf q}\nu}$ with 
the energy $\omega^{\rm ph}_{{\bf q}\nu}$ of a phonon with momentum ${\bf q}$ 
and the coefficients $g^{\rm e-b}$ 
with the electron-phonon matrix elements $g^{\rm e-ph}$:\cite{Giustino2007prl}
\begin{equation} \label{eq:eph-matel}
 g^{\rm e-ph}_{mn\nu}(\bk,\bq)=\langle\psi_{m\bk+\bq}|{\Delta}_{\bq\nu}V_\textup{KS}|\psi_{n\bk}\rangle,
\end{equation}
where $\psi_{n\bk}$ denote Bloch single-particle states and ${\Delta}_{\bq\nu}V_\textup{KS}$ the variation of the self-consistent 
Kohn-Sham potential\cite{hohenbergkohn,kohnsham1965} with respect to a phonon perturbation.
In polar materials, such as TiO$_2$, the electron-phonon coupling 
matrix elements exhibit a $1/| \mathbf{q}|$ singularity 
stemming from the induced electric field generated by the 
finite Born effective charges of the ions.\cite{Froelich1954,Ponce2015}
Such singularity is treated analytically in the following 
according to Refs.~\onlinecite{Verdi2015} and \onlinecite{Sjakste2015}, 
whereby an efficient computational procedure has been established 
by devising an interpolation scheme based on maximally-localized 
Wannier functions\cite{Marzari2012} that accounts for the long-range, singular part of the 
electron-phonon matrix elements. 

The electron self-energy due to electron-plasmon coupling\cite{PhysRevB.94.115208}
can also be evaluated through Eq.~(\ref{eq:sigma}).
In this case, $\omega^{\rm b}_{\bf q}$ coincides with the 
plasmon energy $\omega^{\rm pl}_{\bf q}$
and the dependence on $\nu$ may 
be dropped as we will consider systems with a single plasmon mode. 
The electron-plasmon coupling coefficients take the form:\cite{PhysRevB.94.115208}
  \begin{align}\label{eq:gs}
  g^{\rm e-pl}_{mn}({\bf k},{\bf q})=
  & \left[ \left.\frac{\partial\epsilon({\bf q},\omega)} 
   {\partial\omega}\right|_{\omega^{\rm pl}_{\bf q}} \right]^{-\frac{1}{2}} \nonumber\\ 
& \times {\left(\frac{4\pi} {\Omega_{\rm BZ}}\right)}^{\frac{1}{2}} 
  \frac{1}{|{\bf q}|} \langle \psi_{m{\bf k+q}} |e^{i{\bf q}\cdot{\bf r}} | \psi_{n{\bf k}} \rangle,
  \end{align}
where $\epsilon$ is the electronic 
dielectric function.
As revealed by Eq.~\eqref{eq:gs}, the electron-plasmon coupling matrix elements $g^{\rm e-pl}$
also exhibit an integrable $1/|{\bf q}|$ singularity for vanishing momentum. 
This behaviour is analogous to the Fr\"ohlich electron-phonon coupling matrix elements,  
suggesting that (i) the electron-plasmon interaction is also dominated by the 
coupling with long-wavelength plasmons, and 
(ii) phenomena that are characteristic fingerprints of the coupling  to phonons
(such as, e.g., lifetime effects,\cite{Verdi2015} the renormalization of 
quasiparticle energies,\cite{Logothetidis1992,Giustino2010prl,Antonius2015prl} and the 
formation of photoemission kinks\cite{lanzara/2001} and satellites\cite{Moser2013prl,Baumberger2016,Verdi2017}) 
may result also from the interaction with plasmons, as demonstrated by recent 
theoretical\cite{PhysRevB.94.115208} and experimental\cite{EuO,Jang901} 
studies on highly-doped semiconductors and oxides. 

The numerical evaluation of the electron-plasmon and electron-phonon self-energies via 
Eqs.~\eqref{eq:sigma}-\eqref{eq:gs}
provides the starting point to investigate the combined 
effects of plasmons and phonons on the angle-resolved 
spectral properties of highly doped semiconductors and 
their influence on the formation of satellites.
The spectral function in the diagonal approximation 
for the Fan-Migdal self-energy can be expressed as:{
\begin{align} \label{eq:A}
A_{n{\bf k}} (\omega) = \frac{1}{\pi} \frac{|{{\rm Im}}\,\Sigma^{\rm tot}_{n{\bf k}}(\omega)  |} 
{[\omega - \varepsilon_{n{\bf k}} - {\rm Re}\,\Sigma^{\rm tot}_{n{\bf k}}(\omega)]^2 + [{{\rm Im}}\,\Sigma^{\rm tot}_{n{\bf k}}(\omega)]^2}, 
\end{align} 
where we have defined the total electron self-energy
$\Sigma^{\rm tot}_{n{\bf k}} = \Sigma_{n{\bf k}}^{\rm e-ph} + \Sigma_{n{\bf k}}^{\rm e-pl}$.}
The spectral function is closely related to the photo-electron current\cite{Hedin1998}  
and it thus provides the standard expression for relating theoretical and 
experimental photoemission spectra. However, this approach falls short when it comes 
to photoemission satellites, as it typically overestimates the 
satellite binding energy and its intensity.\cite{Ferdi1996,guzzo/2011}
A more accurate description of the spectroscopic signatures 
of the electron-boson interaction can be obtained from the 
cumulant expansion approach,\cite{Gumhalter/2016/PRB,Sky2015,Kas2014,Nery-2017} 
whereby the spectral function may be rewritten as:\cite{Verdi2017} 
  \begin{align} \label{eq.ourcalc}
  A(\bk,\omega)={\sum}_n \left[ A_{n\bk}^{\rm{QP}}\right. &(\omega)+
  A_{n\bk}^{\rm S1}(\omega) \ast A_{n\bk}^{\rm{QP}}(\omega) \\
  &\left.+ \frac{1}{2} A_{n\bk}^{\rm S1}(\omega) \ast  A_{n\bk}^{\rm S1}(\omega) \ast A_{n\bk}^{\rm QP}(\omega) \right].\nonumber
  \end{align}
Here, $A_{n\bk}^{\rm{QP}}(\omega)$ is obtained by replacing {$\Sigma^{\rm tot}_{n{\bf k}}(\omega)$} by 
{$\Sigma^{\rm tot}_{n{\bf k}}(\ve_{n{\bf k}})$} in Eq.~(\ref{eq:A}), and we defined:
 \begin{equation}\label{eq-spectrum2}
  A_{n\bk}^{\rm S1}(\omega)
  =  \frac{\beta_{n{\bf k}}(\omega) - \!\beta_{n{\bf k}}(\varepsilon_{n{\bf k}}) -
  \!(\omega-\varepsilon_{n{\bf k}})\!\left.
  \displaystyle\frac{\partial \beta_{n{\bf k}}}{\partial \omega}
    \right|_{\varepsilon_{n{\bf k}}}}
  {(\omega-\varepsilon_{n{\bf k}})^2},
  \end{equation}
with 
{$\beta_{n{\bf k}}(\omega) = {\pi^{-1}}{\rm Im}\,\Sigma_{n{\bf k}}^{\rm tot}(\varepsilon_{n{\bf k}}-\omega)
\theta(\omega)$.}
The second (third) term in Eq.~(\ref{eq.ourcalc}) accounts for the spectral 
features arising from the simultaneous excitation of a hole and one (two) 
bosons, whereas $A^{\rm{QP}}$ accounts for the spectral structures 
corresponding to quasiparticle excitations.\cite{Verdi2017}

\section{The electron-plasmon self-energy}\label{sec:imp} 

%-------------------------------------------
 \begin{figure}[t]
    \includegraphics[width=0.48\textwidth]{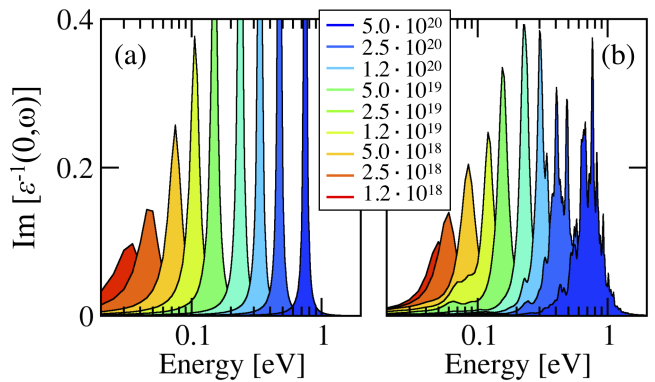} 
   \caption{(a) Loss function of doped silicon for a range of 
   experimentally accessible carrier concentrations (in cm$^{-3}$) obtained from 
   Eq.~\eqref{eq:hegmodel}. (b) Same as in panel (a), obtained from the RPA. 
  }
 \label{fig:loss1}
 \end{figure}

While the electron-phonon self-energy for polar materials may be obtained 
from  well-established first-principles codes,\cite{Ponce2016a}
calculations of the electron-plasmon self-energy for doped compounds have 
only recently become accessible.\cite{PhysRevB.94.115208}
Below, we define a procedure for the 
evaluation of the electron-plasmon self-energy that is 
suitable for first-principles calculations.
As the inspection of Eq.~\eqref{eq:gs} reveals, the computation of the electron-plasmon 
matrix elements entails the evaluation of the dielectric function in the presence of doping, 
which requires ultra-fine ${\bf k}$-point sampling of the Brillouin zone.
This task, due to its high-computational cost, can be accomplished via
direct evaluation of the Kohn-Sham states at each ${\bf k}$-point only 
for the simplest semiconductors, such as silicon, and  
it is thus important to 
define a computational procedure that circumvents this difficulty. 

As a first step, we obtain a simplified expression for the dielectric function of 
doped semiconductors and oxides. 
We start by approximating the momentum-energy dispersion relations through a  parabolic model 
for doping-induced charge carriers close to the band edges, whereas we retain the 
full (first-principles) dispersion for the remaining electrons.
This approximation is well justified for doped semiconductors, 
whereby the Fermi energy typically differs by less than 0.1~eV 
from the band edges. 
The total dielectric function in the presence 
of doping $\epsilon^{\rm D}$ can thus be expressed as:
\begin{align}\label{eq:hegmodel}
\epsilon^{\rm D}({\bf q},\omega) = 
1 - v(\bf q) \left[ \chi^{\rm I} ({\bf q},\omega) + \chi^{\rm HEG} ({\bf q},\omega) \right]
\end{align}
where $v ({\bf q}) = 4\pi/{\bf q}^2 $ is the Coulomb interaction.
$\chi^{\rm I}$ is the polarizability of the insulating (undoped) 
system evaluated within the random-phase approximation (RPA).
$\chi^{\rm HEG}$ is the RPA polarizability of a 
homogeneous electron gas with electron density coinciding with the concentration of extrinsic 
carriers and mass $m_{\rm b}$ corresponding to the isotropic effective mass of the semiconductor 
conduction band. 
$\chi^{\rm HEG}$ can be evaluated analytically\cite{Mahan2000} 
and in the static limit it reduces to: 
\begin{align}\label{eq:rpaheg}
\chi^{\rm HEG}(\text{q},0) = {\frac{{q}^2_{\rm TF}}{8\pi} \left[ -1+ 
\frac{x^2-1}{2x} {\rm ln}\left| \frac{1+x}{1-x}\right| \right], }
\end{align}
where $x=q/2 k_{\rm F}$, $q^{\rm TF}$ is the Thomas-Fermi momentum, and $k_{\rm F}$ the Fermi momentum. 
 
We validate this approximation for the dielectric function of 
doped-semiconductors by comparing in Fig.~\ref{fig:loss1} the loss 
function $L(\omega)={\rm Im}[\epsilon^{\rm D}({\bf q}=0,\omega)]^{-1}$
of doped silicon obtained from Eq.~\eqref{eq:hegmodel} (a) with the results of 
a full calculation based on the RPA (b) 
at ${\bf q}=0$ for $n$-type doping concentrations in the range $10^{18}$-$10^{21}$~cm$^{-3}$. 
In both panels of Fig.~\ref{fig:loss1}, the peaks in the loss function indicate the excitations of 
carrier plasmons.
Electron-hole transitions (not shown) are excited only at energies larger 
than the direct fundamental gap of silicon ($E_{\rm g}=3.3$~eV)
 and they are effectively left  unchanged by the extrinsic doping at these carrier concentrations.
The energy of the plasmon peaks in the RPA and 
its dependence on the carriers density is consistent with the plasma energy
$\omega^{\rm pl} = {(4\pi n /m_{\rm b}\epsilon_\infty)}^{\frac{1}{2}}$.
As illustrated in Fig.~\ref{fig:loss1}, the approximate description of 
doping through Eq.~\eqref{eq:hegmodel} yields a loss function in 
excellent agreement with the RPA.  
The emergence of carrier plasmon peaks is reproduced well and the plasmon energy and intensity are 
in good agreement with the RPA ones even though, for doping larger than $10^{20}$~cm$^{-3}$, 
the intensity of the plasmon peak is somewhat overestimated. 
Overall, the comparison reported in Fig.~\ref{fig:loss1} 
validates the use of Eq.~\eqref{eq:hegmodel} for describing the 
dielectric function of doped semiconductors for a wide range of 
doping levels. 
Generally we expect this description to hold whenever
the energy separation between the conduction band and 
other bands is much larger than the plasmon energy.

Having defined an approximate procedure to estimate $\epsilon^{\rm D}$, 
we discuss in the following how to compute the electron-plasmon coupling matrix elements. 
For the sake of numerical stability, it is desirable to circumvent the explicit evaluation of 
the derivative term in Eq.~\eqref{eq:gs}.
As illustrated in the supplemental material of Ref.~\onlinecite{PhysRevB.94.115208}, 
for $|{\bf q}|< q_{\rm c}$, where $q_{\rm c} = k_{\rm F}\left[
(1+\omega^{\rm pl}/\varepsilon_{\rm F})^{1/2}-1\right]$ is the 
critical momentum that marks the onset of Landau damping 
and $\varepsilon_{\rm F}$ the Fermi energy,
the derivative term in Eq.~\eqref{eq:gs} can be expressed as: 
\begin{align}\label{eq-derev}
\left[\left.\frac{\partial\epsilon}{\partial\omega}
\right|_{\omega_{\bf q}^{\rm pl} }\right]^{-1}  = -\frac{\omega^{\rm pl}_{\bf q}}{2} 
\left[
\epsilon^{\rm D}({\bf q},0)^{-1} -
\epsilon^{\rm I}({\bf q},0)^{-1}\right].
\end{align}
where $\epsilon^{\rm I}$ is the dielectric function of 
the undoped (insulating) system.
This expression for the dielectric function relies on the assumptions that 
the energy of interband electron-hole transitions is much larger than 
the plasmon energy, and that plasmons and phonons may be treated independently, 
that is, possible phenomena arising from plasmon-phonon coupling, 
such as plasmon-phonon polaritons, are neglected. 
Noting that plasmons may only be excited in a narrow region of crystal 
momenta close to ${\bf q}=0$, we further introduce the approximation
$\langle \psi_{m{\bf k+q}} |e^{i{\bf q}\cdot{\bf r}} | \psi_{n{\bf k}} \rangle = \delta_{nm}$, 
and we obtain an explicit expression for the electron-plasmon coupling matrix elements: 
\begin{align}
\left| g^{\rm e-pl}_{mn}({\bf k}, {\bf q}) \right|^2 = 
{\frac{2\pi\delta_{nm} \omega^{\rm pl}_{\bf q}}{\Omega_{\rm BZ}} }
\frac{\left[
\epsilon^{\rm D}({\bf q},0)^{-1} -
\epsilon^{\rm I}({\bf q},0)^{-1}\right] }
{|{\bf q} |^2 }.\nonumber
\end{align}
where $\epsilon^{\rm D}$ is given by Eqs.~\eqref{eq:hegmodel}-\eqref{eq:rpaheg}.
{
This expression may be further simplified by noting that, 
for $q < q_{\rm c}$, it is a good approximation to consider 
$\epsilon^{\rm I}({\bf q},0) \simeq \epsilon_\infty$, where $\epsilon_\infty$ is the high-frequency dielectric 
constant which can be obtained from first-principles calculations of the RPA dielectric function 
{in the pristine system}.
The final expression for the electron-plasmon matrix elements can be rewritten as: 
\begin{align}\label{eq:g2plex}
\left| g^{\rm e-pl}_{mn}({\bf k}, {\bf q}) \right|^2 = \delta_{nm} 
\frac{v({\bf q}) \omega^{\rm pl}_{\bf q}}{2\Omega_{\rm BZ} }
{\left[ \frac{1}{\epsilon_\infty - \epsilon^{\rm HEG} ({\bf q}) +1   }
-\frac{1}{\epsilon_\infty} \right]}
\end{align}
where $\epsilon^{\rm HEG} ({\bf q}) = 1 - v ({\bf q}) \chi^{\rm HEG}({\bf q},0)$ 
is the static Lindhard dielectric function.\cite{Mahan2000}
The advantage of this procedure is that the matrix 
elements $g^{\rm e-pl}$ are expressed in terms of quantities available from first-principles 
calculations of undoped compounds, whereas explicit calculations 
in the presence of doping are avoided.
}

We determine the electronic and lattice-dynamical properties of TiO$_2$ from density
functional theory (DFT)  and density functional perturbation theory (DFPT) 
calculations within the generalized gradient approximation\cite{PBE} 
as implemented in the Quantum ESPRESSO package\cite{Giannozzi2017}. 
Only valence electrons are treated explicitly, including the semicore $3s$ and $3p$ 
states of Ti, whereas core electrons are accounted for through Troullier-Martins norm-conserving 
pseudopotentials.\footnote{Available at:
https://github.com/mmdg-oxford/papers/tree/master/Verdi-NCOMMS-2017/pseudo} 
Convergence is ensured by using a 200~Ry kinetic energy cutoff 
and a $6\times6\times6$ Monkhorst-Pack mesh.
The DFT single-particle eigenvalues, the phonon dynamical matrices and 
the electron-phonon matrix elements are first obtained on a homogeneous 
$4\times4\times4$ {Brillouin-zone} grid. 
In order to compute the 
electron-phonon self-energy, the electronic and phononic bands 
as well as the electron-phonon matrix elements are then interpolated 
on a dense random $\mathbf{q}$-point mesh with 168,914 points with a denser sampling of
the region close to $\Gamma$ according to a Cauchy distribution of width 0.01. 
The interpolation is performed as in Ref.~\onlinecite{Verdi2017} 
using maximally-localized Wannier functions within the EPW code~\cite{Ponce2016a} 
through an internal call to the Wannier90 library~\cite{Mostofi2008}. 
The electron-plasmon self-energy has been implemented in the EPW code~\cite{Ponce2016a} 
by combining Eqs.~\eqref{eq:sigma} and \eqref{eq:g2plex} and by taking 
advantage of the Wannier interpolation of the electronic energies, and it has been 
computed using the same random grid. 
We describe doping within the rigid-band approximation, whereby extrinsic 
carriers are accounted for by means of a rigid shift of the Fermi energy. 
Charge neutrality is maintained through the addition of a homogeneous 
positively charged background. 

%-------------------------------------------
\begin{figure}[t]
 \includegraphics[width=\columnwidth]{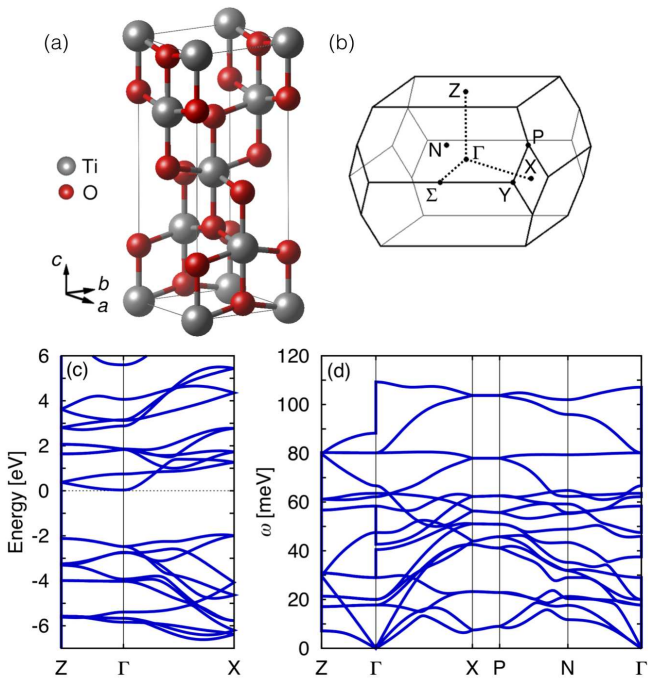} 
 \caption{ \label{fig:band-ph} 
(a) Atomistic model of the unit cell of anatase \tio.  
(b) Brillouin zone and high-symmetry lines. 
(c) Calculated electronic band structure along the directions parallel ($\Gamma$X) and 
perpendicular ($\Gamma$Z) to the basal plane of the Brillouin zone. 
(d) Phonon dispersions computed along high-symmetry lines in the Brillouin zone.
 }
\end{figure}
%-------------------------------------------

\section{Hybrid plasmon-phonon satellites in photoemission}\label{sec:spec}
%
%-------------------------------------------
\begin{figure}[t]
\includegraphics[width=0.48\textwidth]{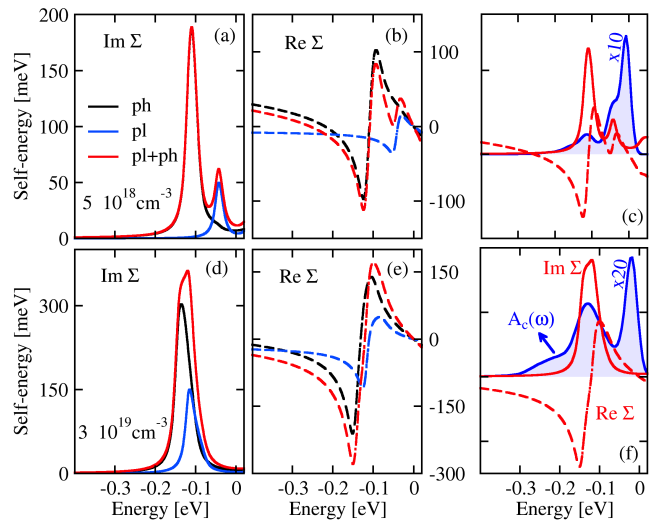} 
\caption{(a)-(b) Imaginary (a) and real (b) part of the electron self-energy 
due to the electron-phonon (ph) and electron-plasmon (pl) interaction for \tio at 
a doping concentration of 5$\cdot10^{18}$~cm$^{-3}$. 
(d)-(e) Same as panels (a) and (b) for 3$\cdot10^{19}$~cm$^{-3}$. 
(c), (f) Comparison between the total self-energy and 
the cumulant spectral function due to the combined effect of plasmons and phonons 
at $\Gamma$ for 5$\cdot10^{18}$ and 3$\cdot10^{19}$~cm$^{-3}$, respectively. }
\label{fig:sigma}
 \end{figure}
%-------------------------------------------

In the following, we employ the formalism presented in Secs.~\ref{sec:th} and \ref{sec:imp} 
to investigate the formation of plasmon and phonon satellites in anatase \tio. 
The atomistic model for the unit cell of \tio is shown in Fig.~\ref{fig:band-ph}~(a), 
whereas the Brillouin zone and the high-symmetry points are illustrated in (b). 

The electron and phonon band structures calculated within DFT and DFPT 
are displayed in Fig.~\ref{fig:band-ph}~(c) and (d), respectively. 
The low-energy conduction bands
derive from the strongly localized Ti $3d$ states. The conduction-band 
bottom lies at the $\Gamma$ point and it is formed by a single band with strongly 
anisotropic character in the directions perpendicular and parallel to the 
$c$ axis. 
The phonon dispersion relations in Fig.~\ref{fig:band-ph}~(b) show large LO-TO 
splittings for the three infrared-active modes. The highest-energy phonon is the $E_u$ mode at 
109~meV, which has been identified as the main source of Fr\"ohlich-type electron-phonon 
coupling in anatase\cite{Moser2013prl,Verdi2017}. 

We proceed to investigate the coupling of electrons to plasmons and 
phonons through the calculation of the electron self-energy. 
The real and imaginary parts of $\Sigma^{\rm tot}_{n{\bf k}}$ for the 
conduction band at the $\Gamma$ point are 
illustrated in Fig.~\ref{fig:sigma} for carrier concentrations of 
5$\cdot10^{18}$ [(a) and (b)] and 3$\cdot10^{19}$~cm$^{-3}$ [(d) and (e)], and 
they are compared to the real and imaginary parts of the electron-plasmon and electron-phonon  
self-energies.  
The total spectral function due to the combined effect of plasmons
and phonons is obtained from  
the cumulant expansion through Eqs.~\eqref{eq.ourcalc}-\eqref{eq-spectrum2} using 
$\Sigma^{\rm tot}_{n{\bf k}}$  as a seed. 
In panels (c) and (f) the cumulant spectral function 
at $\Gamma$ is reported alongside with the total self-energy, illustrating that the 
satellite features in the spectral functions 
occur at energies corresponding to peaks in the imaginary part of the self-energy.
All energies are relative to the Fermi level. Our calculations are performed using 
the same carrier densities as the ARPES experiment of Ref.~\onlinecite{Moser2013prl}, 
as determined from the measured three-dimensional Fermi surfaces. 
The bulk-sensitivity of these measurements
has been proven via the inspection of the Fermi surface, which
demonstrates the three-dimensional character of the band dispersion\cite{Moser2013prl}
as further validated by the agreement with
first-principles calculation.\cite{Verdi2017}
For $n=$5$\cdot10^{18}$~cm$^{-3}$ (3$\cdot10^{19}$~cm$^{-3}$), the 
concentration of extrinsic carriers is described through 
a 13~meV (40~meV) shift of the Fermi level above the conduction-band bottom. 
Temperature effects are accounted for by considering Fermi-Dirac and 
Bose-Einstein occupation factors at 20~K in Eq.~\eqref{eq:sigma} and 
by multiplying the spectral functions by the Fermi-Dirac distribution function 
$f(\omega,T)=(1+e^{\omega/k_{\rm B}T})^{-1}$, where $k_{{\rm B}}$ is the Boltzmann constant.
To provide a picture of the photoemission process in closer agreement with experiment, 
we further account for finite resolution effects in energy and momentum through 
the convolution with Gaussian functions of widths 25~meV and 0.015~\AA$^{-1}$, respectively. 
All spectral function calculations presented in the following 
are based on the cumulant expansion approach with the inclusion of
finite resolution effects.

%-------------------------------------------
 \begin{figure*}[t]
    \includegraphics[width=0.95\textwidth]{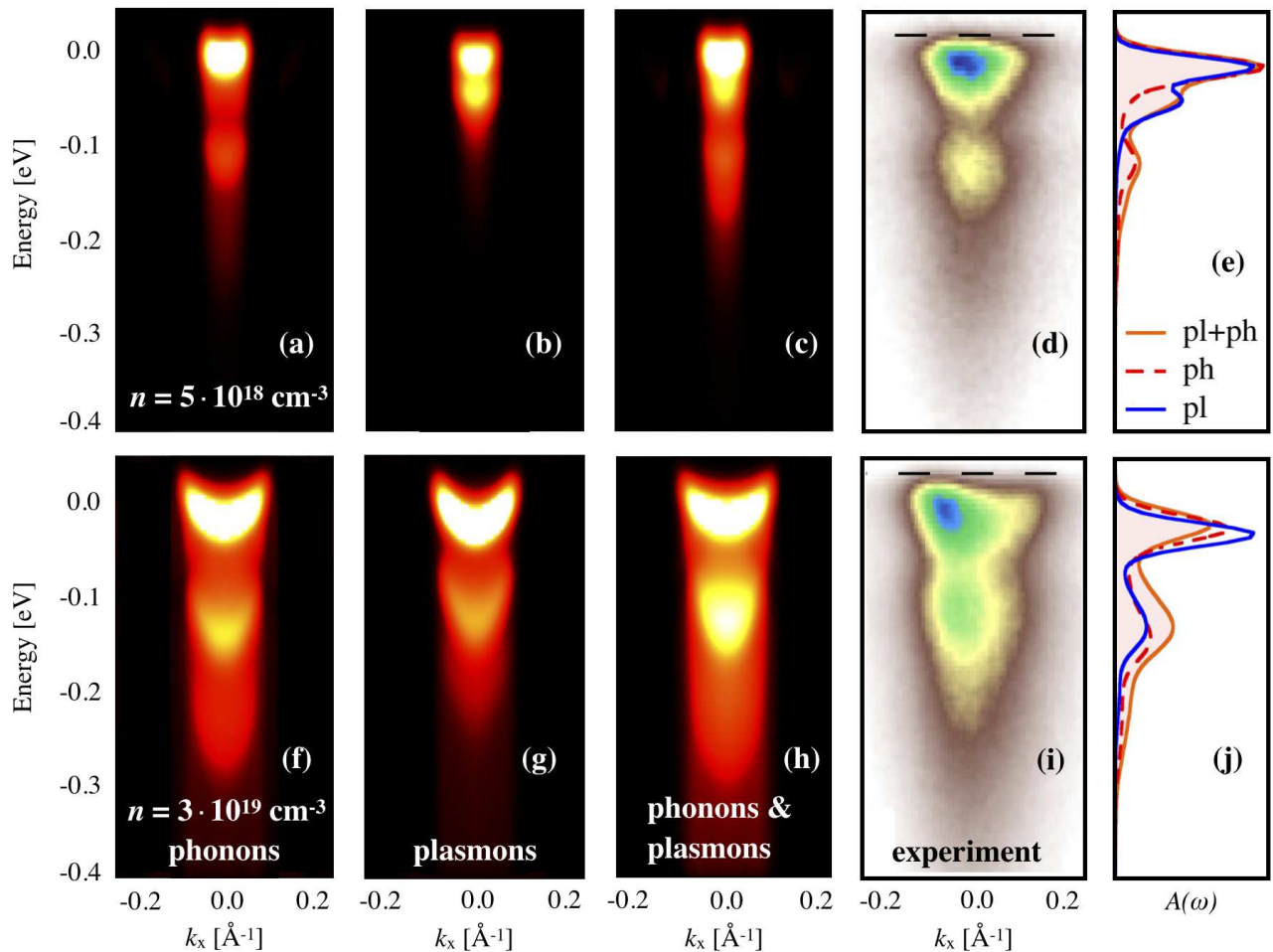} 
   \caption{
(a)-(d) Spectral function of \tio obtained from the cumulant expansion 
approach by accounting for the effects of electronic coupling 
to (a) phonons, (b) plasmons, and (c) both at a doping concentration of 5$\cdot10^{18}$~cm$^{-3}$. 
(d) Angle-resolved photoemission spectrum of \tio, adapted from Ref.~\onlinecite{Moser2013prl}. 
(e) Spectral function at $\Gamma$ due to 
plasmons (pl), phonons (ph), and their combined effect (pl+ph) from the cumulant expansion.
(f)-(j) Same as above for $n=$3$\cdot10^{19}$~cm$^{-3}$. }
 \label{fig:spec2}
 \end{figure*}

Figure~\ref{fig:spec2} illustrates the calculated 
spectral function of \tio at the 
conduction-band bottom for $n=$5$\cdot10^{18}$ [(a)-(e)] and 3$\cdot10^{19}$~cm$^{-3}$ 
[(g)-(j)], along the same line in reciprocal space as the ARPES experiment\cite{Moser2013prl}. 
To disentangle the contribution of different mechanisms to the total spectrum, 
we further report cumulant spectral functions obtained separately from the 
electron-phonon [(a) and (f)]  and electron-plasmon [(b) and (g)] self-energies.
For comparison, we report in Fig.~\ref{fig:spec2}~(d) and (i) 
ARPES measurements of the conduction-band bottom of \tio from  
Ref.~\onlinecite{Moser2013prl}, and in panels (e) and (j) the 
cumulant spectral function at the $\Gamma$ point. 

All the spectra in Fig.~\ref{fig:spec2} exhibit sharp spectral features 
at low binding energies ($\omega<50$~meV) that follow a parabolic dispersion. 
These structures, 
which stem from the excitation of quasiparticle states at the conduction-band bottom, 
result from the shift of the Fermi level inside the conduction band. 
At binding energies between 40 and 300~meV, the spectral functions are 
characterized by the emergence of additional features which may not be 
attributed to quasi-particle excitations, revealing doping-dependent 
effects of electron-boson interactions.  
The case of electron-phonon interaction shown in Fig.~\ref{fig:spec2}~(a) and (f) 
has been thoroughly discussed elsewhere.\cite{Verdi2017} In brief, at both doping values
the spectral function exhibits well-defined satellite features 
separated from the bottom of the conduction band by an energy compatible with the 
energy of the $E_u$ LO phonon of \tio, 109~meV.
Additional broadened low-intensity features are observed 
at binding energies between 150 and 300~meV, which may be attributed to 
two-phonon processes.

The electron-plasmon spectral functions in Fig.~\ref{fig:spec2}~(b) and \ref{fig:spec2}~(g) 
are also characterized by distinct satellite features which, 
in this case, may be ascribed to the excitation of a photo-hole 
and a plasmon. 
These results indicate that, at sufficiently high dopant concentrations, 
the strong coupling to plasmons may lead to the emergence of distinctive 
satellite structures in the angle-resolved spectral function of \tio. 
These features are reminiscent of the plasmonic polaron 
bands observed in the valence bands of (undoped) 
semiconductors\cite{caruso/2015,caruso/2015/2} at 
binding energies of the order of 10-15~eV and in model systems.\cite{Caruso2016EPJB}
In doped TiO$_2$, however, the satellite binding energy is comparable to 
the case of polaronic coupling:
for $n=$5$\cdot10^{18}$~cm$^{-3}$, the satellite is separated from the 
conduction band by $32$~meV, whereas for 
$n=$3$\cdot10^{19}$~cm$^{-3}$ we obtain an energy difference of $79$~meV.
The remarkable increase of satellite binding energies
arises from the dependence of the plasmon energy $\omega^{\rm pl}$ 
on the carrier concentration $n$.
The phonon energy, on the other hand,  can be expected to exhibit 
a rather weak dependence ($<5-8$~meV) on the 
doping concentration which can primarily be attributed to the 
emergence of non-adiabatic effects beyond the Born-Oppenheimer 
approximation.\cite{Caruso/PRL/2017} Non-adiabatic corrections 
to the phonon energy have been neglected here. 

The total spectral function due to both plasmons and phonons is shown in Fig.~\ref{fig:spec2}~(c) and \ref{fig:spec2}~(h) and 
its inspection reveals two different scenarios for the interplay of plasmons and phonons:
(i) For $n=$5$\cdot10^{18}$~cm$^{-3}$, the energy of the plasmon satellite $\omega^{\rm pl}=32$~meV 
is comparable to the experimental resolution of 25~meV and significantly 
smaller than the $E_u$ LO phonon energy $\omega^{\rm ph} = 109$~meV. 
Correspondingly, the intensity of the plasmon satellite merges 
with the quasiparticle state, whereas the electron-phonon 
interaction yields an isolated polaronic satellite. 
(ii) For the largest carrier density, on the other hand, 
the plasmon energy $\omega^{\rm pl}=79$~meV differs from the 
phonon energy approximately by the energy resolution. 
The spectral signatures of plasmon and phonon satellites are thus
degenerate in energy, leading to the formation of a hybrid 
plasmon-phonon satellite with enhanced intensity as compared 
with the electron-phonon spectral function of Fig.~\ref{fig:spec2}~(f).

To quantify the contribution of plasmons to the total 
electron-boson interaction strength, we estimate the 
parameter $\lambda_{\bf k} = -\partial\Sigma_{\bf k} / 
\partial\omega |_{\omega = \varepsilon_{\rm F}}$, 
which is related to the enhancement of the band effective mass $m^*$  via the 
relation $m^* = m_{\rm b} (1+\lambda)$, where $m_{\rm b}$ denotes the {\it bare} mass ($\Sigma=0$)\cite{grimvall1981electron} 
and $\lambda$ is the Fermi-surface average of $\lambda_{\bf k}$.
We estimate $\lambda_{\bf k}$ by taking the derivative of 
the self-energy at the Fermi level along the $\Gamma$X direction. 
For the electron-phonon (electron-plasmon) coupling we obtain 
$\lambda^{\rm ph} = 0.68,0.66$ ($\lambda^{\rm pl}=0.43,0.32$) 
for $n=$5$\cdot10^{18}$ and 3$\cdot10^{19}$~cm$^{-3}$.
Combining the contribution of plasmons and phonons we obtain 
$\lambda^{\rm pl-ph} = 1.09,1.00$ for the two doping levels, suggesting that the 
electron-plasmon interaction contributes by approximately 
30-40\% to the total electron-boson coupling strength.
This result also suggests that the coupling of electrons to 
plasmons is stronger at smaller doping concentrations, provided that the 
doping level is high enough to sustain carrier plasmon excitations in the system, 
and that their characteristic energy is large enough to detect the 
effects of the electron-plasmon interaction within the experimental 
resolution. 

The inspection of the quasiparticle weight $Z = 1/(1+\lambda)$ 
may provide further indications of the effects of plasmons 
on the electronic properties of doped semiconductors and oxides. 
Earlier experimental studies\cite{Moser2013prl} on \tio have reported 
a quasiparticle weight $Z^{\rm exp} = 0.36$ for a doping concentration of 
3$\cdot10^{19}$~cm$^{-3}$. 
If only electron-phonon interactions are accounted for, 
first-principles calculations yield $Z^{\rm ph}= 0.59$ 
which overestimates significantly the experimental value, 
indicating that phonons contribute to only a fraction of the 
total electron-boson coupling strength. 
Remarkably, by combining the effects of plasmons and phonons, 
we obtain a quasiparticle weight  $Z^{\rm pl-ph} = 0.5$, which  
improves the agreement between theory and experiment. 
Overall, these results indicate that, while phonons provide the 
dominant contribution to the formation of polaronic satellites in 
\tio, electron-phonon coupling alone yields only a fraction of the 
total electron-boson coupling strength necessary to explain the 
small quasiparticle weight measured experimentally. 
Electron-plasmon coupling, as well as other mechanisms 
such as, e.g., scattering with electron-hole pairs and impurities,
may prove important to explain the residual discrepancy between
measured and calculated quasiparticle weights in highly-doped 
semiconductors and oxides. 
The measured ratio between the 
intensity of quasiparticle and satellite peaks, 
and thus the quasiparticle weight, might further be 
influenced by the extrinsic effects. 
The photon energy employed to resolve polaronic features
in TiO$_2$ ($h\nu\simeq85$~eV) 
is typically smaller than that previously employed for the observation of
valence satellites in undoped semiconductors ($h\nu\sim800~$eV) \cite{guzzo/2011}.
Photons thus have a smaller penetration depth in the sample and,
correspondingly, the photoelectrons are expected to exhibit weaker
signatures of extrinsic effects. To quantitatively establish the
influence of extrinsic effects on the experimental photoemission spectrum
of doped TiO$_2$, however, a more in-depth investigation based
on quantum-mechanical simulations would be required, e.g., by following the approach
presented in Ref.~\onlinecite{Guzzo2012}.

Recently, a combination of ARPES experiments and first-principles 
calculations has been employed to investigate the electronic properties of 
highly-doped EuO samples as a function of carrier concentration.\cite{EuO}
In addition to quasiparticle peaks, the ARPES spectra of EuO exhibit clear 
signatures of polaronic satellites at low-carrier densities 
($n\simeq10^{18}$cm$^{-3}$), which, in analogy to the case of \tio, 
at higher doping concentrations are accompanied by carrier
plasmon satellites. 
In combination with the present study, these
results suggest that the emergence of 
carrier plasmon satellites may constitute a general 
feature of highly-doped semiconductors and oxides.

\section{Hot-carrier relaxation due to plasmons and phonons}\label{sec:LW}

\begin{figure}[t]
\includegraphics[width=0.35\textwidth]{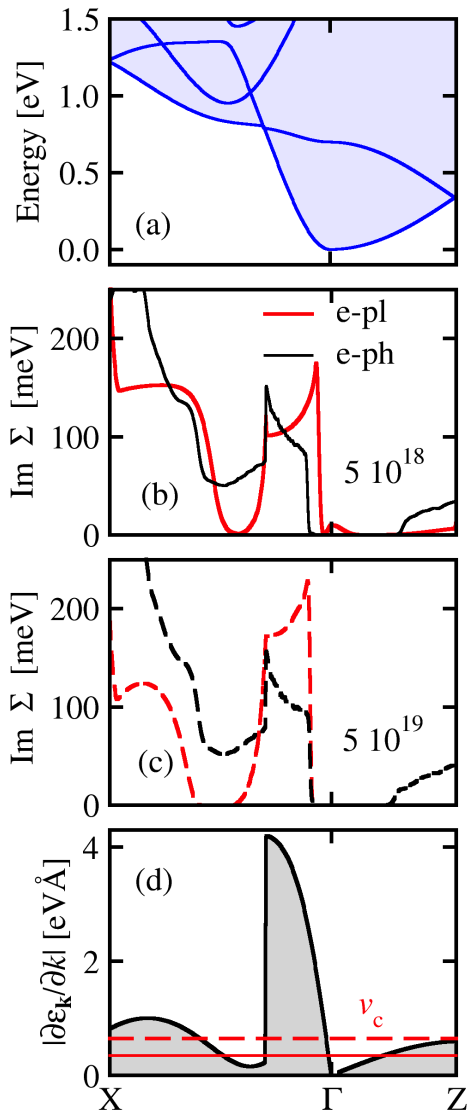}
\caption{(a) Lowest energy conduction bands of TiO$_2$ obtained from DFT. 
(b)-(c) Electron linewidth due to the electron-phonon and electron-plasmon interaction 
for a carrier concentration of (b) 5$\cdot10^{18}$~cm$^{-3}$ and (c) 3$\cdot10^{19}$~cm$^{-3}$. 
(d) Band velocity $v_{n{\bf k}}$ for the conduction band and critical velocity $v_{\rm c}$
for the onset of electron-plasmon scattering for 5$\cdot10^{18}$~cm$^{-3}$ (continuous)  and  3$\cdot10^{19}$~cm$^{-3}$ (dashed).
}
\label{fig:LW}
\end{figure}

To quantify the effects of phonons and plasmons on the dynamics of 
excited carriers in the conduction band of TiO$_2$, illustrated in 
Fig.~\ref{fig:LW}~(a), we evaluate the low-temperature electron linewidths due to 
the electron-phonon and electron-plasmon interaction. 
The electron linewidths $\Gamma_{n{\bf k}}$ are obtained from the 
self-energy via  $\Gamma_{n{\bf k}}={\rm Im} \Sigma_{n{\bf k}} (\varepsilon_{n{\bf k}})$ 
and, owing to their relation to the scattering time {$\tau_{n{\bf k}} = 1/2\Gamma_{n{\bf k}}$, }
they provide direct information pertaining the characteristic timescale of 
electron-boson scattering and their influence on the relaxation process for 
excited carriers.

In Fig.~\ref{fig:LW} we compare the electron linewidths due to  
electron-phonon and electron-plasmon coupling at carrier concentrations of 
5$\cdot10^{18}$~cm$^{-3}$ (b) and 3$\cdot10^{19}$~cm$^{-3}$ (c)  
for crystal momenta along the X-$\Gamma$-Z high-symmetry path.
In both cases, the linewidth vanishes in the vicinity of the 
conduction-band minimum since both plasmon and phonon emission 
are forbidden by energy conservation. 
For carrier energies smaller than 0.5~eV, we obtain electron-phonon 
linewidths of the order of 100~meV, corresponding to relaxation times
of about 3~fs. 
Electron-plasmon scattering, on the other hand, 
yields linewidths around 10~meV (20~meV) 
for $n=$5$\cdot10^{18}$~(3$\cdot10^{19}$~cm$^{-3}$), corresponding 
to relaxation times of 33~fs (16~fs). 
%$\hbar=0.65821195$eV$\cdot$fs$\rightarrow\tau=..$ 
Overall, the electron-plasmon linewidths are significantly smaller than their 
electron-phonon counterpart suggesting that, at these doping concentrations, 
carrier relaxation is dominated by electron-phonon scattering.
However, at variance with the electron-phonon linewidths which eventually  
decrease with the increase of doping concentration owing to 
the screening of the electron-phonon interaction, the 
electron-plasmon linewidths increase with the carrier 
density,\cite{PhysRevB.94.115208} suggesting that, at 
higher doping values, electron-plasmon and electron-phonon 
scattering may contribute in a similar way to the relaxation 
of excited carriers.

Interestingly, the electron-plasmon linewidths vanish 
for several momenta in the Brillouin zone, a phenomenon that we attribute 
to the lack of phase space for the scattering between plasmons and excited carriers. 
In fact, an excited carrier with energy $\epsilon_{n{\bf k}}$ and crystal momentum ${\bf k}$ 
can emit a plasmon with momentum ${\bf q}$ and energy $\omega^{\rm pl}_{\bf q}$ 
only if there exists an empty final state with momentum ${\bf k-q}$ 
and energy $\epsilon_{n{\bf k-q}} = \epsilon_{n{\bf k}} -\omega^{\rm pl}_{\bf q}$.
Since plasmons can be excited only for ${q < q_{\rm c}}$, 
this condition can be obeyed only if $| v_{n{\bf k}} |  > \omega^{\rm pl}_{\bf q} / q_{\rm c}$, 
where $v_{n{\bf k}} = \partial \epsilon_{n{\bf k}} /\partial {\bf k} $ is the band velocity. 
In other words, an excited carrier can undergo electron-plasmon scattering only 
if its band velocity is larger than the critical velocity $v_c =  \omega^{\rm pl}_{\bf q} / q_{\rm c}$. 
{Based on a similar argument, one may also expect the Fr\"ohlich coupling to LO phonons 
to be stronger for electronic states with a higher band velocity owing to the 
increased phase space for electron-phonon scattering. } 
To further validate this argument, we report in Fig.~\ref{fig:LW}(d) the 
electron velocity along X-$\Gamma$-Z and we indicate the critical 
velocity $v_c$ by horizontal lines for both doping concentrations. 
The comparison with Fig.~\ref{fig:LW}(b) and (c) confirms that the 
electron-plasmon linewidths vanish whenever $|v_{n{\bf k}} | < v_c$. 
Thus, the relation $|v_{n{\bf k}} | > v_c$ provides a new 
necessary condition, of general validity, for the emergence of 
electron-plasmon scattering. 
From this condition, it also follows that electrons 
in very dispersive bands (that is, with small effective masses)
will be subject to electron-plasmon scattering to a larger extent 
owing to the larger phase-space available for these processes.

\section{Conclusions}\label{sec:conc}

We investigated the interplay of plasmons and phonons 
on the spectral properties of $n$-doped anatase \tio.
In order to describe the coupling to carrier plasmons (resulting from the 
extrinsic carriers introduced via doping), we have derived 
and implemented a numerical approach to compute
the electron self-energy due to the electron-plasmon 
interaction from first principles. 
This formalism has been applied to investigate the simultaneous influence of 
plasmons and phonons on (i) the formation of polaronic satellites in the 
ARPES spectra of highly-doped \tio, and 
(ii) the characteristic relaxation timescales for excited carriers 
in the conduction band. 

Our study reveals that electron-plasmon coupling may underpin the 
formation of photoemission satellites at energy scales of the order of 
30-80~meV, in close analogy with the well-known scenario of polaronic coupling. 
At variance with phonons, however, the plasmon-satellite energy 
depends strongly on doping via the carrier concentration. 
First-principles calculations of the ARPES spectral function indicate that 
at low carrier concentrations, when the plasmon energy approaches the 
experimental resolution, satellite and quasiparticle bands become
indistinguishable, leading to the formation of a single, broad spectral feature
close to the Fermi energy. 
At higher concentrations, the plasmon and phonon energies become comparable and 
their satellites merge into a more intense spectral feature.
The explicit treatment of electron-plasmon interaction 
is important to improve the agreement with experiments
for the quasiparticle weight $Z$, indicating that a 
significant fraction of the total electron-boson coupling strength 
can be attributed to plasmons. 
Finally, the computation of relaxation times suggests 
that, at the carrier concentrations considered 
here, electron-plasmon scattering occurs 
at a much lower rate as compared to electron-phonon 
scattering, indicating that phonons dominate the  
relaxation of excited carriers in the conduction band 
of \tio. 

In conclusion, these results suggest that the interplay of electrons, 
plasmons, and phonons underpin a complex scenario of many-body interaction in \tio, 
whereby the simultaneous coupling to different bosonic modes depends pronouncedly
on the concentration of extrinsic carriers and influences pervasively
photoemission satellites, linewidths, and quasiparticle weights. 
These aspects emphasize the importance of explicitly accounting for the effects of 
both phonons and plasmons in future studies of highly-doped semiconductors and oxides. 
Accounting for these phenomena through first-principles 
approaches  may provide a valuable tool to unravel the fundamental 
quantum-mechanical processes that underpin the formation of satellites 
in photoemission spectroscopy and the relaxation of excited carriers. 

\acknowledgments 
Discussions with Marco Grioni are gratefully acknowledged. 
The research leading to these results has received funding from the Leverhulme Trust (Grant RL-2012-001),
the UK Engineering and Physical Sciences Research Council (grant No. EP/M020517/1), and the Graphene
Flagship (EU FP7 grant no. 604391). The authors acknowledge the use of the University of Oxford Advanced
Research Computing (ARC) facility (http://dx.doi.org/10.5281/zenodo.22558). We acknowledge
PRACE for awarding us access to Cartesius at SURFsara, Netherlands (DECI-13); Abel at UiO, Norway
(DECI-14); and MareNostrum at BSC-CNS, Spain (PRACE-15).

\end{document}